\documentclass[aps,twocolumn,preprintnumbers,amsmath,amssymb,nofootinbib,superscriptaddress,notitlepage]{revtex4-1}

\usepackage{epsfig}

\usepackage[utf8]{inputenc}

\begin{document}

\title{The $\Omega(2012)$ as a hadronic molecule}

	\author{M. Pavon Valderrama}\email{mpavon@buaa.edu.cn}
\affiliation{School of Physics and Nuclear Energy Engineering, \\
International Research Center for Nuclei and Particles in the Cosmos and \\
Beijing Key Laboratory of Advanced Nuclear Materials and Physics, \\
Beihang University, Beijing 100191, China} 

\date{\today}


\begin{abstract}
  \rule{0ex}{3ex}
  Recently the Belle collaboration has discovered
  a narrow $S=-3$ baryon, the $\Omega(2012)$.
  We explore the possibility that the $\Omega(2012)$
  is a $\Xi(1530)\,\bar K$ molecule, where the binding mechanism is
  the coupled channel dynamics with the  $\Omega\,\eta$ channel.
  The characteristic signature of a molecular $\Omega(2012)$ will be its decay
  into the three body channel $\Xi \pi \bar{K}$, for which we
  expect a partial decay width of $2-3\,{\rm MeV}$.
  The partial decay width into the $\Xi \bar{K}$ channel should lie
  in the range of $1-11\,{\rm MeV}$, a figure compatible
  with experiment and which we have deduced from the assumption
  that the coupling involved in this decay is of natural size.
  For comparison purposes the decay of a purely compact $\Omega(2012)$ into
  the $\Xi \bar{K}$ and $\Xi \pi \bar{K}$ channels is
  of the same order of magnitude as and one order of magnitude smaller
  than in the molecular scenario, respectively.
  This comparison indicates that the current experimental information
  is insufficient to distinguish between a compact and a molecular
  $\Omega(2012)$ and further experiments will be required
  to determine its nature.
  A molecular $\Omega(2012)$ will also imply the existence of two- and
  three-body molecular partners.
  The two-body partners comprise two $\Lambda$ hyperons located at $1740$
  and $1950\,{\rm MeV}$ respectively, the first of which might correspond
  to the $\Lambda(1800)$ while the second to
  the $\Lambda(2000)$ or the $\Lambda(2050)$.
  The three-body partners include a $\Xi(1530) K\bar{K}$ and
  a $\Xi(1530) \eta \bar{K}$ molecule,
  with masses of $M = 2385-2445\,{\rm MeV}$
  and $M = 2434-2503\,{\rm MeV}$ respectively.
  We might be tempted to identify the first with the $\Xi(2370)$ and
  the latter with the $\Omega(2470)$ listed in the PDG.
\end{abstract}

\maketitle

The discovery of the $\Omega(1673)$
baryon~\cite{Barnes:1964pd} confirmed
the SU(3)-flavour symmetry of Gell-Mann~\cite{GellMann:1962xb}
and Ne'eman~\cite{Ne'eman:1961cd} as the ordering
principle of baryon spectroscopy.
After five decades
the $\Omega(1673)$ remains to be the only four star $S=-3$ baryon resonance.
Only other three $S=-3$ baryons are listed
in the PDG~\cite{Patrignani:2016xqp}, the three star
$\Omega(2250)$ and the two-star $\Omega(2380)$ and $\Omega(2470)$.
Recently the Belle collaboration discovered a new addition to the family,
a narrow $\Omega$ baryon with a mass $M = 2012.4 \pm 0.7 \pm 0.6$ and a
width $\Gamma = 6.4^{+2.5}_{-2.0} \pm 1.6$~\cite{Yelton:2018mag}.

The most prosaic explanation for the nature of the new $\Omega(2012)$ baryon
is that of a decuplet $\frac{3}{2}^{-}$ compact state,
as indicated by the Belle collaboration itself~\cite{Yelton:2018mag}
and predicted for instance in the Isgur-Karl model
at $2020\,{\rm MeV}$~\cite{Isgur:1978xj}.
Recent theoretical works have explored this idea further
from the point of view of the chiral quark model~\cite{Xiao:2018pwe},
QCD sum rules~\cite{Aliev:2018syi} and SU(3) flavour
symmetry~\cite{Polyakov:2018mow},
in all cases suggesting the quantum numbers $J^P = \frac{3}{2}^{-}$.
As a matter of fact the existence of a $\frac{3}{2}^-$ $\Omega$ baryon
in the $2.0-2.1\,{\rm GeV}$ region was already expected
from the quark model~\cite{Chao:1980em}, large $N_c$~\cite{Goity:2003ab},
the Skyrme model~\cite{Oh:2007cr} and lattice QCD~\cite{Engel:2013ig}.

Here we consider the possibility that the $\Omega(2012)$ is molecular
instead of fundamental, as proposed in Ref.~\cite{Polyakov:2018mow}.
Molecular hadrons are a prolific theoretical concept, which indeed
explain the properties of a few hadrons that do no fit
into the quark model, see Refs.~\cite{Guo:2017jvc,Ali:2017jda,Olsen:2017bmm}
for recent reviews.
But there are a few drawbacks to this idea too:
for instance there is no uniform description of hadronic molecules,
a few of the approaches are phenomenological and lack a clear
connection with QCD and what constitute a molecular
state is sometimes a nebulous concept.
In this regard it is important to work in detail, if possible,
how the theoretical description of a specific molecule relates to other
molecular candidates, whether their internal dynamics can be related
to the low energy manifestations of QCD (for instance, chiral symmetry)
and how the predictions for fundamental and compound hadrons differ.

Here by a compound $\Omega(2012)$ we refer to a hadron
in the line of the $\Lambda(1405)$ or $D_{s0}(2317)$,
which are suspected
to be $N \bar{K}$~\cite{Jido:2003cb,Magas:2005vu,Hyodo:2007jq}
and $D K$~\cite{Guo:2006fu,Guo:2009ct} molecules respectively
(the list of possible molecular candidates is long and growing,
see Ref.~\cite{Guo:2017jvc} for instance).
Analogously to the two previous examples, the mechanism responsible for the
binding of the $\Omega(2012)$ will be the Weinberg-Tomozawa (WT)
interaction between a baryon and a pseudo Nambu-Goldstone boson.
The difference lies in the requirement of coupled channel dynamics:
besides the $\Xi(1530) \bar{K}$, which is a natural molecular
explanation for the $\Omega(2012)$~\cite{Polyakov:2018mow},
the $\Omega \eta$ channel will also be involved.
In fact the WT interaction in the $\Xi(1530) \bar{K}$-$\Omega \eta$
system is~\cite{Sarkar:2004jh,Si-Qi:2016gmh}
\begin{eqnarray}
  V = -\frac{\omega + \omega'}{2 f^2}
  \begin{pmatrix}
      0 & 3 \\
      3 & 0 
  \end{pmatrix}
  \, ,
\end{eqnarray}
with $\omega$ and $\omega'$ the incoming and outgoing energies of the
pseudo Nambu-Goldstone bosons and where we are taking
the normalization $f = f_{\pi} = 132\,{\rm MeV}$.
This interaction is attractive and particularly strong in the eigenchannel
\begin{eqnarray}
  \frac{1}{\sqrt{2}}\left[ | \Xi(1530) \bar{K} \rangle +
    | \Omega \eta \rangle \right] \, ,
\end{eqnarray}
where its strength indeed matches that of the $N\bar{K}$ interaction
generating the $\Lambda(1405)$ and surpasses that of the $D K$
potential that gives rise to the $D_{s0}(2317)$.
The $\Xi(1530) \bar{K}$-$\Omega \eta$ interaction is known to be able
to generate a pole~\cite{Sarkar:2004jh,Si-Qi:2016gmh}.
The difference with the $N\bar{K}$ and $D K$ cases is however
that we are dealing with a coupled channel problem:
the interaction happens in the non-diagonal channels where
there is a gap of $192\,{\rm MeV}$ between the channels.
This requires to check whether such a bound state can arise naturally
from the leading order WT interaction without forcing
an unnaturally large cut-off.
For that we first introduce a cut-off by multiplying the potential
by a regulator function depending on each of the external momenta
\begin{eqnarray}
  V \to V\,g(\frac{p'}{\Lambda})\,g(\frac{p}{\Lambda}) \, ,
\end{eqnarray}
with $p$ and $p'$ the incoming and outgoing momenta of the mesons,
where we simply choose a Gaussian regulator $g(x) = e^{-x^{2n}}$ with $n = 2$.
If we include this potential into a dynamical equation\footnote{Here we will
  follow the formalism of Refs.~\cite{Valderrama:2018knt,Valderrama:2018sap}
  for the two- and three-body integral equations (latter we will show
  three-body results).
  For the two-body system, the eigenvalue equation we are using is
  \begin{displaymath}
    \phi_i(p) = \int \frac{d^3 q}{(2\pi)^3} \frac{m_i}{\omega_i(q)}
    \frac{\langle p | V_{ij} | q \rangle
      \phi_j(q)}{E_i - \frac{q^2}{2 M_i} - \epsilon_i(q)} \, ,
  \end{displaymath}
  where $i=1,2$ refers to the $\Xi(1530) \bar{K}$ and $\Omega \eta$ channels, 
  $m_i = m_K, m_{\eta}$ and $M_i = M_{\Xi^*}, M_{\Omega}$ are the pseudo
  Goldstone-Boson and Baryon masses,
  $E_i$ the energy with respect to the $i$-channel threshold,
  $\omega_i(q) = \sqrt{m_i^2 + q^2}$ and $\epsilon_i(q) = \omega_i(q) - m_i$.
  That is, we will treat the baryons as non-relativistic and
  the pseudo Nambu-Goldstone bosons as relativistic.
  The dynamical equation in the two-body sector will be equivalent
  to the Kadyshevsky equation~\cite{Kadyshevsky:1967rs},
  except for the detail of the non-relativistic baryon.
},
it turns out that for generating a pole at $2012\,{\rm MeV}$ a cut-off of
$\Lambda = 721\,{\rm MeV}$ is needed.
This is to be compared with the $\Lambda(1405)$ and $D_{s0}(2317)$ poles,
which require a cut-off of $\Lambda = 571\,{\rm MeV}$ and
$\Lambda = 823\,{\rm MeV}$ respectively~\footnote{
If we consider $f_K$ instead of $f_{\pi}$ for the strength of
the Weinberg-Tomozawa, the cut-off at which the $\Omega(2012)$ binds
will be $890\,{\rm MeV}$ instead, to be compared with $762\,{\rm MeV}$
for the $\Lambda(1405)$ and $1042\,{\rm MeV}$ for the $D_{s0}(2317)$.
Other subleading order corrections will modify the cut-off too,
likely within the range of the previous estimates.
}.
The bottom-line is that the cut-off for reproducing the $\Omega(2012)$ pole
is sensible.
From this we can argue that the existence of a
$\Xi(1530) \bar{K}$-$\Omega \eta$ molecule is consistent
with the natural expectations derived from other similar molecules.
Indeed this molecule has been previously predicted to be
at $2141 - 38 i\,{\rm MeV}$~\cite{Sarkar:2004jh}
and $1786\,{\rm MeV}$~\cite{Si-Qi:2016gmh}, where the different location
will correspond a lower and higher cut-off in the regularization
we are using here, respectively.
In the absence of experimental information about the location of
a candidate $\Omega^*$ pole, the previous predictions
are completely plausible.

\begin{figure}[ttt]
\begin{center}
  \includegraphics[width=3.5cm]{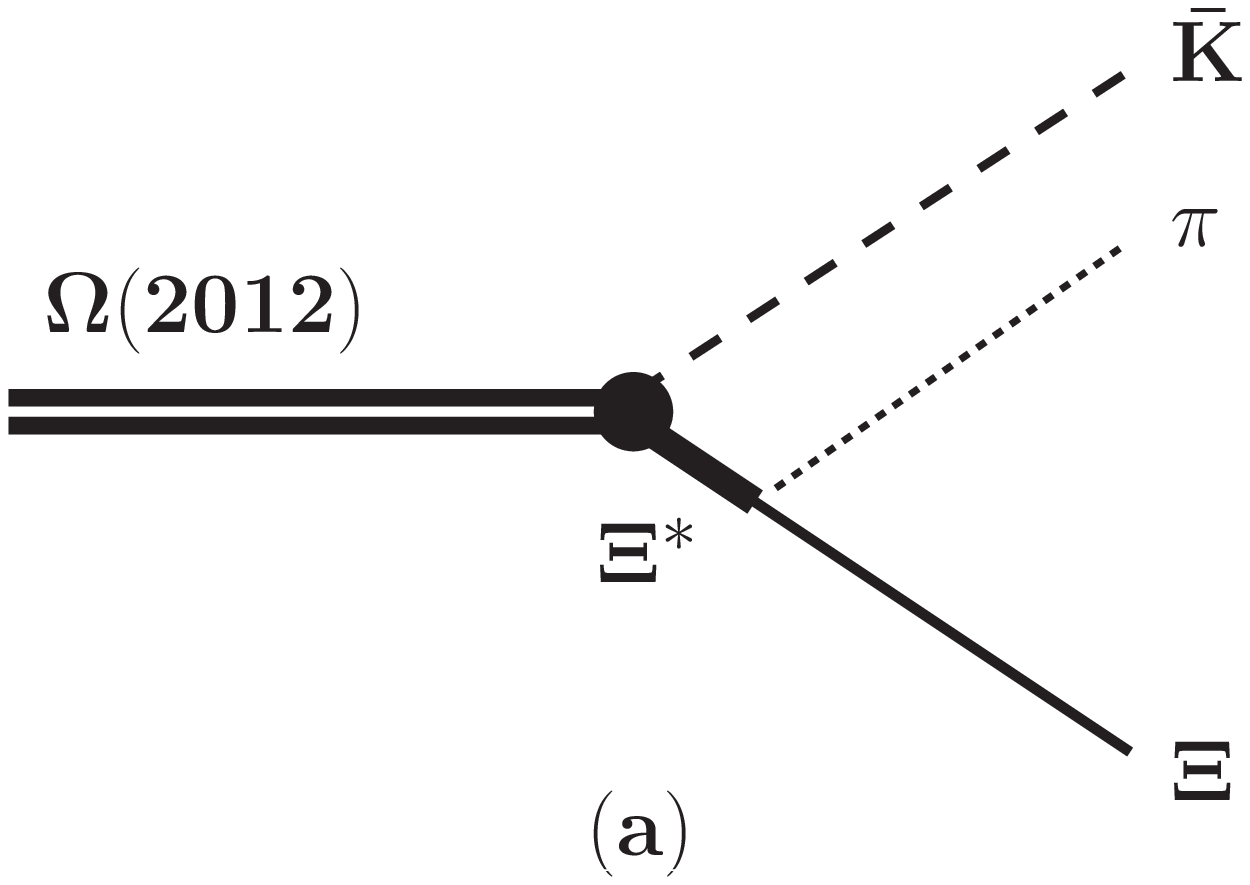}
  \includegraphics[width=4.5cm]{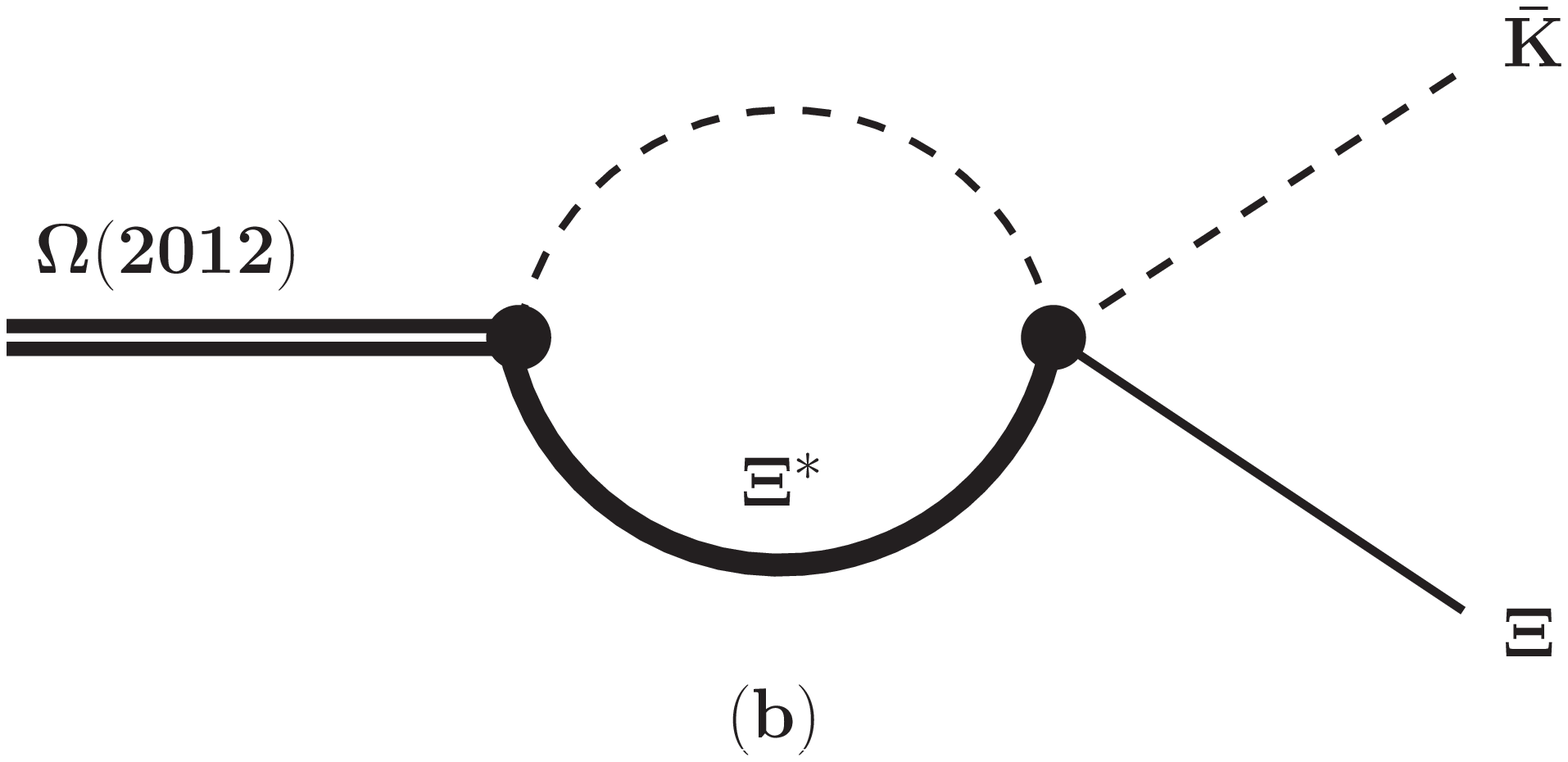}
\end{center}
\caption{
  Feynman diagrams involved in the decays of a molecular $\Omega(2012)$:
  (a) represents the three body decay into $\Xi \pi \bar{K}$,
  which is the characteristic signature of a molecular $\Omega(2012)$,
  while (b) represents the two body decay into $\Xi \bar{K}$,
  which happens via a short-range operator.
}
\label{fig:diagrams}
\end{figure}

Yet this by itself is not enough to decide whether
the $\Omega(2012)$ is really compatible with the molecular hypothesis.
For this it is also necessary to compute its decays,
if possible with theoretical uncertainties.
We consider the decays into $\Xi \pi \bar{K}$ and $\Xi \bar{K}$
via the mechanisms shown in Fig.~\ref{fig:diagrams}.
For the theoretical uncertainties we will modify the WT interaction as follows
\begin{eqnarray}
  V = -\frac{\omega + \omega'}{2 f^2}\,c(\Lambda)\,
  g(\frac{p'}{\Lambda}) g(\frac{p}{\Lambda})\,
  \begin{pmatrix}
      0 & 3 \\
      3 & 0 
  \end{pmatrix}
  \, ,
\end{eqnarray}
that is, we will consider that the coupling runs with the cut-off,
instead of being fixed. We then determine $c(\Lambda)$ from
the condition of reproducing the $\Omega(2012)$ pole.
The cut-off will vary in the $\Lambda = 0.5-1.0\,{\rm GeV}$ window,
which comprises the {\it privileged} cut-off for which
$c(\Lambda) = 1$, i.e. $\Lambda = 721\,{\rm MeV}$.
The cut-off window is wide enough as to accommodate the changes of
this privileged cut-off owing to subleading order corrections
to the baryon-meson interaction.
Finally we will interpret the variation of the results
with the cut-off as the uncertainty of our calculations.

The most defining feature of a molecular $\Omega(2012)$
will be the three body decay $\Omega(2012) \to \Xi \pi \bar{K}$,
see Fig.~\ref{fig:diagrams}.
The partial width of this process can indeed be directly deduced from:
(i) the decay $\Xi(1530) \to \Xi \pi$
where $\Gamma \simeq 9-10\,{\rm MeV}$~\cite{Patrignani:2016xqp}
and (ii) the wave function of the $\Omega(2012)$,
which determines the coupling to the $\Xi(1530) \bar{K}$ channel.
Concrete calculations yield a partial decay width of
\begin{eqnarray}
  \Gamma(\Omega^* \to \Xi \pi \bar{K}) \simeq 2-3\,{\rm MeV} \, ,
\end{eqnarray}
where $\Omega^*$ refers to the $\Omega(2012)$.
This width is smaller than that of the $\Xi(1530)$ in the same channel
as a consequence of the existence of an $\Omega \eta$ component
in the wave function and that the antikaon takes out momentum of
the pion, further reducing the partial decay width.
The details of the calculation are analogous to those of
the $X(3872) \to D\bar{D} \pi$ partial decay width
in the molecular description of the $X(3872)$,
which is deduced from the $D^* \to D \pi$ amplitude~\cite{Guo:2014hqa}.
The only difference with the work of Ref.~\cite{Guo:2014hqa} is that
here we do not include $\Xi \bar{K}$ rescattering effects,
as the $\Xi \bar{K}$ WT term vanishes.
Subleading order corrections to the $\Xi(1530)\bar{K}$-$\Omega \eta$
potential will have a moderate impact in the $\Xi \pi \bar{K}$
decay width: there will be a diagonal term in the potential
that will change the probability of the $\Xi(1530)\bar{K}$
component in the wave function.
However taking into account the large uncertainty of the lowest order result,
the inclusion of subleading effects is probably not justified.
Finally if the $\Omega(2012)$ is a compact state, the three body
$\Xi \pi \bar{K}$ decay width is probably of the order of
$50-100\,{\rm KeV}$.
This figure, which we have deduced from phase space, the angular
momentum of the final three body state and the size of the decay
coupling as estimated from naive dimensional analysis~\cite{Manohar:1983md},
is remarkably smaller than in the molecular scenario.

Yet the decay that is experimentally known is $\Omega(2012) \to \Xi \bar{K}$,
which for a molecular $\Omega(2012)$ happen via the mechanism
depicted in the right panel of Fig.~\ref{fig:diagrams}.
The short-range potential involved in this decay
is a $\Xi(1530) \bar{K} \to \Xi \bar{K}$ vertex of the type
\begin{eqnarray}
  \langle \Xi \bar{K}(\vec{p}') | V | \Xi^* \bar{K}(\vec{p}) \rangle =
  C_D\,\vec{S}\,\cdot \vec{q}\,\,\vec{\Sigma}\, \cdot \vec{q} \, ,
\end{eqnarray}
where $\Xi^*$ refers to the $\Xi(1530)$, $\vec{\Sigma}$ stands for
the spin-$\frac{3}{2}$ matrices of the $\Xi^*$,
$\vec{S}$ are spin-$\frac{3}{2}$ to -$\frac{1}{2}$ transition
matrices and $\vec{q} = \vec{p}\,' - \vec{p}$
is the momentum transfer~\footnote{
  The matrix element can be equivalently written in terms of
  $\vec{S}\,\cdot \vec{q}\,\vec{\sigma} \cdot \vec{q}$,
  where $\sigma$ are the Pauli matrices for the final $\Xi$ baryon.
  Explicit expressions for the $\vec{S}$ matrices
  can be found in Refs.~\cite{Lu:2017dvm,Haidenbauer:2017sws}.
  We have chosen the momentum transfer instead of the antikaon initial
  and final momenta for convenience: they are equivalent however,
  as these dependencies can be interchanged by means of
  the equations of motion.}.
The size of the coupling can be estimated
from naive dimensional analysis~\cite{Manohar:1983md}
\begin{eqnarray}
  C_D \sim \frac{1}{f^2 \Lambda_{\chi}} \, ,
\end{eqnarray}
with $f$ the pion decay constant and $\Lambda_{\chi} \sim 1\,{\rm GeV}$
the chiral symmetry breaking scale.
This gives us
\begin{eqnarray}
  \Gamma(\Omega^* \to \Xi \bar{K}) \sim 2-11\,{\rm MeV} \quad (1-5\,{\rm MeV})
  \, ,
\end{eqnarray}
for $f = f_{\pi}$ ($f = f_{K} \simeq 160\,{\rm MeV}$).
The details of the calculation are not presented here, but they are again
analogous to the decay of the theoretical $X(4012)$ molecule
into $D\bar{D}$, which were presented in Ref.~\cite{Albaladejo:2015dsa}
(the only difference is that the decay mechanism in this case
is a contact-range operator, instead of one pion exchange).

A few things are worth noticing: the diagram leading to the $\Xi \bar{K}$
decay is linearly divergent, which partly explains the spread of
the width in the $\Lambda = 0.5-1.0\,{\rm GeV}$ range.
This is not crucial for the previous estimation, which is there to give
a sense of whether the molecular hypothesis is compatible with experiment.
But from the point of view of an effective field theory description
the interpretation is as follows~\cite{Valderrama:2014vra,Valderrama:2016koj}:
(i) the coupling $C_D$ is actually a {\it running coupling}
with scales as the inverse of the cut-off and
(ii) as a consequence of this scaling the  previous contribution
is enhanced with respect to naive dimensional analysis.
Notice too that there are also long-range contributions to the two-body decay:
if we consider the original three body decay, there is the possibility
that the pion rescatters with the antikaon with the latter absorption of
the pion by the final $\Xi$ cascade, i.e. a triangle diagram.
However this contribution is strongly divergent and cannot be evaluated
without the inclusion of higher order counterterms.

\begin{table}[ttt]
  \begin{tabular}{|c|c|c|c|c|}
    \hline
    Nature & $J^P$ & $\Delta^*$ & $\Gamma(\Delta^* \to N\pi)$ &
    $\Gamma(\Omega^* \to \Xi \bar{K})$ \\
    \hline \hline
    $\Xi^* \bar{K}$-$\Omega \eta$ & $\frac{3}{2}^{-}$ & - & - &
    $2-11$ ($1-5$)  \\
    \hline \hline
    $sss$ & $\frac{1}{2}^-$ & $\Delta(1620)$ & $25-49$
    & $30-60$ \\
    \hline
    $sss$ & $\frac{3}{2}^+$ & $\Delta(1600)$ & $16-81$
    & $13-65$ \\
    \hline
    $sss$ & $\frac{3}{2}^-$ & $\Delta(1700)$ & $20-60$
    & $5-14$ \\
    \hline
  \end{tabular}
  \caption{
    Estimations of the partial decay width (in $\rm MeV$) of
    the $\Omega(2012)$ into $\Xi \bar{K}$ in the molecular
    and quark state scenarios.
    The notation $\Omega^*$ and $\Xi^*$ refer to the $\Omega(2012)$
    and $\Xi(1530)$ respectively.
    For a molecular $\Omega^*$ the decay is mediated via a contact-range
    $\Xi^* \bar{K} - \Xi \bar{K}$ interaction, which size is estimated
    from naive dimensional analysis.
    For a compact $\Omega^*$ the calculation of this partial decay width
    can be obtained from SU(3)-flavour symmetry, but it requires
    in the first place to identify the $\Omega^*$ as
    the member of a known multiplet.
  }
  \label{tab:decays}
\end{table}

This is to be compared with a compact, non-molecular $\Omega(2012)$,
in which case the $\Omega(2012)$ will belong to a decuplet
with a $\Delta$ isobar partner.
As a consequence, the $\Omega^* \to \Xi \bar{K}$ decay width can be related
to the $\Delta^* \to N \pi$ decay width.
The concrete relation can be worked out easily
from $SU(3)$-flavour symmetry\cite{Samios:1974tw}
\begin{eqnarray}
  \Gamma(\Omega^* \to \Xi \bar{K}) =
  2\,\frac{M_{\Delta^*}}{M_{{\Omega}^*}}\,
  {\left( \frac{p_K}{p_{\pi}} \right)}^{2l+1}\,
  \Gamma(\Delta^* \to N \pi) \, , \nonumber \\
\end{eqnarray}
where the factor $2$ is a consequence of the different $SU(3)$ Clebsch-Gordan
coefficients, $M_{\Omega^*}$, $M_{\Delta^*}$ are the masses of
the baryons involved and $p_K$ and $p_{\pi}$ are the decay momenta of
the antikaon and pion respectively.
If we assume the quantum numbers of the $\Omega(2012)$ to be
$J^P = \frac{1}{2}^{-}$, $\frac{3}{2}^{+}$ and $\frac{3}{2}^{-}$,
then it will belong to the same decuplet as the $\Delta(1620)$,
$\Delta(1600)$ and $\Delta(1700)$ respectively.
Now we can make predictions for the partial decay width to $\Xi \bar{K}$,
which can be consulted in Table~\ref{tab:decays}.
From this the only identification that is within the experimental limits
for the width is the $J^P = \frac{3}{2}^{-}$, for which
the partial decay width is $5-14\,{\rm MeV}$.
This range is on average broader than, but still compatible with,
the experimental value.
From this comparison the compact and molecular hadron hypotheses
are equivalent in what regards the total decay width.
Besides, the theoretical calculation of the $\Omega^* \to \Xi \bar{K}$
decay width is subjected to large uncertainties independently of
whether the $\Omega(2012)$ is a three quark or a molecular state.
For a three quark $\Omega^*$ it is worth noticing that flavour symmetry is
not expected to work as well for excited baryons as it does
for the fundamental ones, particularly regarding masses
(yet from Ref.~\cite{Samios:1974tw}, where flavour symmetry relations were
successfully applied to excited baryons, SU(3)-flavour seems to work
well enough for our purposes).
For a molecular $\Omega^*$ the estimate for the $\Xi \bar{K}$
decay width hinges on dimensional analysis estimations,
which can be perfectly off by a numerical factor of order one
(explaining, for instance, the smaller partial decay
width obtained in Ref.~\cite{Lin:2018nqd}).
The admixture of the compact and molecular components cannot be ruled out
either, as this might be already happening to its $\Delta$ isobar partner:
the $\frac{3}{2}^{-}$ $\Delta(1700)$ can indeed be reproduced from the chiral
interaction among the $\Delta \pi$-$\Sigma K$-$\Delta \eta$
baryon-meson channels~\cite{Sarkar:2004jh}.
Other example is the $\frac{1}{2}^-$ octet, which can be viewed as
dynamically generated~\cite{Ramos:2002xh} comprising the 
$N(1535)$, $\Lambda(1670)$, $\Sigma(1620)$ and $\Xi(1620)$,
or as a standard SU(3) octet~\cite{Guzey:2005vz} comprising
$N(1535)$, $\Lambda(1670)$, $\Sigma(1570)$ and $\Xi(1620)$,
though in this case there is a mismatch about which
$\Sigma^*$ baryon completes the multiplet.

\begin{table*}[tth]
  \begin{tabular}{|c|c|c|c|c|c|}
    \hline
    Molecule & $I$($J^P$) & $C_{\rm WT}$ & $B$ & $M$ & Candidate \\
    \hline \hline
    $\Xi^* \bar{K}$-$\Omega \eta$ & $0$($\frac{3}{2}^{-}$) &
    $\begin{pmatrix}
      0 & -3 \\
      -3 & 0 
    \end{pmatrix}$
    & 16  & 2012  & $\Omega(2012)$ \\
    \hline 
    $\Xi K$ & $0$($\frac{1}{2}^{-}$)  & -3 & $69-73$ & $1739-1743$ &
    $\Lambda(1800)$ \\ 
    \hline
    $\Xi^* K$ & $0$($\frac{3}{2}^{-}$) & -3 & $77-80$ & $1948-1951$ &
    $\Lambda(2000)$, $\Lambda(2050)$ \\ 
    \hline \hline
    $\Xi^* \bar{K} K$-$\Omega \eta K$ &
    $\frac{1}{2}$($\frac{3}{2}^+$) & - & $78-138$ &
    $2385-2445$ & $\Xi(2370)$, $\Xi(2500)$ \\
    \hline
    $\Xi^* \bar{K} \eta$-$\Omega \eta \eta$ & $0$($\frac{3}{2}^+$)
    & - & $57-126$ & $2434-2503$ & $\Omega(2470)$\\
    \hline
  \end{tabular}
  \caption{
    Predicted two- and three-body molecular partners of the $\Omega(2012)$
    as a $\Xi^* \bar{K}$-$\Omega \eta$ bound state.
    We indicate their particle content, their isospin, spin and parity,
    the relative strength of the WT term
    (for the two-body case), the binding energy and mass ($\rm MeV$),
    plus the prospective candidates among experimentally known baryons
    in the PDG~\cite{Patrignani:2016xqp}.
    The relative strength of the WT interaction is defined such that
    $V = C_{WT}\,(\omega + \omega')/2 f^2\,c(\Lambda)\,
    g(p'/\Lambda)\,g(p/\Lambda)$, where we always use the same regulator.
    The bands in the binding energies and masses are a consequence of
    the cut-off variation in the range $\Lambda = 0.5-1.0\,{\rm GeV}$ and
    are interpreted as the theoretical uncertainty of the calculations.
    The binding energy of the three body bound states is calculated
    with respect to the three body thresholds $\Xi^* \bar{K} K$ and
    $\Xi^* \bar{K} \eta$, respectively.
  }
  \label{tab:partners}
\end{table*}

A molecular $\Omega(2012)$ also implies the existence of partner states
that share the same binding mechanism.
The $\Xi K$ and $\Xi^* K$ WT terms are strong in the $I=0$ channel
\begin{eqnarray}
  V = -3\,\frac{\omega + \omega'}{2 f^2}\,
  c(\Lambda)\,g(\frac{p'}{\Lambda})\,g(\frac{p}{\Lambda}) \, ,
\end{eqnarray}
which probably implies the existence of $\Xi K$ and $\Xi^* K$ bound states
with the quantum numbers of a $\Lambda$ baryon ($I=0, S=-1$).
We find two states located at $1740$ and $1950\,{\rm MeV}$
with quantum numbers $\frac{1}{2}^{-}$ and $\frac{3}{2}^{-}$ respectively,
see Table \ref{tab:partners} for details.
The previous calculations assume that the same cut-off can be used
for the $\Xi$ and $\Xi^*$ cascades.
They also ignore the $\Sigma \pi$ and $\Sigma^* \pi$ channels: they are expected
to move the location of the poles, giving them a finite width in the process.
If the experience with the $\Lambda(1405)$ is of any help, where the location
of the pole moves from $1427\,{\rm MeV}$ to $1428 - 17 i\,{\rm MeV}$
after the inclusion of the $\Sigma \pi$ channel~\cite{Hyodo:2007jq},
we will expect moderate correction at most in the location of
the previous two $\Lambda$ baryons.

In addition, the existence of both $\Xi^* {\bar K}$ and $\Xi^* K$ molecules
suggest the existence of a $\Xi^* K {\bar K} $ bound state,
which can be computed for instance with the formalism of
Refs.~\cite{Valderrama:2018knt,Valderrama:2018sap}.
For this type of three body calculation we have to take into account
the $\Xi^* {\bar K}$ and $\Xi^* K$ WT interaction in the $I=1$ channel,
which is one third of that in the $I=0$ case.
The strength of the $K {\bar K}$ interaction in the isoscalar channel
is determined from the condition of reproducing the $f_0(980)$ pole,
which is thought to be molecular~\cite{Baru:2003qq},
while in the isovector channel
we set it to one third of the strength in the isoscalar channel
as expected for a WT term~\footnote{Notice that in the isovector channel
  the $a_0(980)$ is also expected to have a large $K \bar{K}$ molecular
  component~\cite{Baru:2003qq}, but probably not so large as to determine
  the isovector $K\bar{K}$ interaction from it. If taken into account as
  in the isoscalar case, it will make the three body states a bit more bound.}.
From this  the position of the $\Xi^* K \bar{K}$ three-body bound state
is about $2385-2445\,{\rm MeV}$, where the spread reflects
the cut-off variation.
Notice however that we did not include widths in the calculation:
this molecule will have a sizable width from
the $K {\bar K} \to \pi \pi$ transition.
This three body exotic $\Xi^*$ might be identified with the $\Xi(2370)$ or
maybe with the $\Xi(2500)$ that are included
in the PDG~\cite{Patrignani:2016xqp},
but of which little is known.
Other possible three body partner is a
$\Xi^* {\bar K} \eta$-$\Omega\,\eta\,\eta$ bound state,
which binds owing to the coupled channel dynamics of the $\Omega(2012)$.
If we assume the $\eta$ to be non-interacting (which greatly simplify
the calculations) the location of this state is about $2434-2503\,{\rm MeV}$.
We might identify this $\Omega^*$ with the $\Omega(2470)$ of the PDG.
The results for the molecular partners of the $\Omega(2012)$
are summarized in Table \ref{tab:partners}.
However the $\frac{3}{2}^+$ $\Xi^*$ and $\Omega^*$ excited baryons
we predict in Table \ref{tab:partners} are not necessarily
a signature of a molecular $\Omega(2012)$:
the diagonal and non-diagonal WT interactions of a compact $\Omega(2012)$
with the kaon and the $\eta$ are strong enough as to generate
these $\frac{3}{2}^+$ $\Xi^*$ and $\Omega^*$ states.

To summarize, the molecular hypothesis for the $\Omega(2012)$ is
compatible with the experimentally known information about this baryon
and might be able to explain a few $\Lambda^*$, $\Xi^*$ and $\Omega^*$
baryons listed in the PDG.
The comparison of the molecular and compact baryon scenarios indicates
that the total decay width is roughly identical in both cases,
the only difference being that a molecular $\Omega(2012)$
is expected to have a sizable branching ratio
into $\Xi \pi \bar{K}$ of the order of $30-50\%$.
This branching ratio is at least one order of magnitude smaller
for a compact $\Omega(2012)$, which indicates that this is
the experimental quantity to look for if we want to determine
the nature of this baryon (a point which has been stressed
in the recent literature, see
Refs.~\cite{Polyakov:2018mow,Lin:2018nqd,Huang:2018wth,Pavao:2018xub}).
Other defining feature is what partner states are to be expected in each case.
A compact $\Omega(2012)$ belongs to the $\frac{3}{2}^{-}$ decuplet,
which probably comprises the $\Delta(1700)$ and
two other $\Sigma$ and $\Xi$ baryons that have not been detected yet
with masses of $M_{\Sigma} = 1805 \pm 40$ and
$M_{\Xi} = 1910 \pm 40\,{\rm MeV}$ respectively~\cite{Polyakov:2018mow}.
This decuplet pattern is mostly a consequence of SU(3) flavour symmetry
rather than of the nature of the baryons and it is also dynamically
reproduced by chiral interactions~\cite{Sarkar:2004jh}.
Maybe the defining difference is that a molecular $\Omega(2012)$ will
have additional partners that are dictated by the sign
and strength of the Weinberg-Tomozawa interaction.
In the two-body sector (baryon-meson) these partners comprise two $\Lambda$
hyperons, a $J^P = \frac{1}{2}^{-}$ one with a mass of $1740\,{\rm MeV}$
and another with $J^P = \frac{3}{2}^{-}$ at $1950\,{\rm MeV}$,
which might be identified with the $\Lambda(1800)$ and maybe the
$\Lambda(2000)$ or $\Lambda(2050)$ respectively.
In the three-body sector (baryon-meson-meson)
we find a $J^P = \frac{3}{2}^+$ $\Xi(2400)$ cascade and
a $\frac{3}{2}^+$ $\Omega(2470)$ baryon,
which we are tempted to identify with the $\Xi(2370)$ and
$\Omega(2470)$ listed in the PDG.
Finally, we stress that the current experimental information about
the $\Omega(2012)$ is insufficient to distinguish between
the compact and molecular hypotheses. Future experiments,
in particular regarding the $\Omega^* \to \Xi \bar{K} \pi$
partial decay width, will be necessary
to determine its nature.

\section*{Acknowledgments}

This work is partly supported by the National Natural Science Foundation
of China under Grants No.11522539, No. 11735003, the Fundamental
Research Funds for the Central Universities and
the Thousand Talents Plan for Young Professionals.


\begin{thebibliography}{40}%
\makeatletter
\providecommand \@ifxundefined [1]{%
 \@ifx{#1\undefined}
}%
\providecommand \@ifnum [1]{%
 \ifnum #1\expandafter \@firstoftwo
 \else \expandafter \@secondoftwo
 \fi
}%
\providecommand \@ifx [1]{%
 \ifx #1\expandafter \@firstoftwo
 \else \expandafter \@secondoftwo
 \fi
}%
\providecommand \natexlab [1]{#1}%
\providecommand \enquote  [1]{``#1''}%
\providecommand \bibnamefont  [1]{#1}%
\providecommand \bibfnamefont [1]{#1}%
\providecommand \citenamefont [1]{#1}%
\providecommand \href@noop [0]{\@secondoftwo}%
\providecommand \href [0]{\begingroup \@sanitize@url \@href}%
\providecommand \@href[1]{\@@startlink{#1}\@@href}%
\providecommand \@@href[1]{\endgroup#1\@@endlink}%
\providecommand \@sanitize@url [0]{\catcode `\\12\catcode `\$12\catcode
  `\&12\catcode `\#12\catcode `\^12\catcode `\_12\catcode `\%12\relax}%
\providecommand \@@startlink[1]{}%
\providecommand \@@endlink[0]{}%
\providecommand \url  [0]{\begingroup\@sanitize@url \@url }%
\providecommand \@url [1]{\endgroup\@href {#1}{\urlprefix }}%
\providecommand \urlprefix  [0]{URL }%
\providecommand \Eprint [0]{\href }%
\providecommand \doibase [0]{http://dx.doi.org/}%
\providecommand \selectlanguage [0]{\@gobble}%
\providecommand \bibinfo  [0]{\@secondoftwo}%
\providecommand \bibfield  [0]{\@secondoftwo}%
\providecommand \translation [1]{[#1]}%
\providecommand \BibitemOpen [0]{}%
\providecommand \bibitemStop [0]{}%
\providecommand \bibitemNoStop [0]{.\EOS\space}%
\providecommand \EOS [0]{\spacefactor3000\relax}%
\providecommand \BibitemShut  [1]{\csname bibitem#1\endcsname}%
\let\auto@bib@innerbib\@empty
\bibitem [{\citenamefont {Barnes}\ \emph {et~al.}(1964)\citenamefont {Barnes}
  \emph {et~al.}}]{Barnes:1964pd}%
  \BibitemOpen
  \bibfield  {author} {\bibinfo {author} {\bibfnamefont {V.~E.}\ \bibnamefont
  {Barnes}} \emph {et~al.},\ }\href {\doibase 10.1103/PhysRevLett.12.204}
  {\bibfield  {journal} {\bibinfo  {journal} {Phys. Rev. Lett.}\ }\textbf
  {\bibinfo {volume} {12}},\ \bibinfo {pages} {204} (\bibinfo {year}
  {1964})}\BibitemShut {NoStop}%
\bibitem [{\citenamefont {Gell-Mann}(1962)}]{GellMann:1962xb}%
  \BibitemOpen
  \bibfield  {author} {\bibinfo {author} {\bibfnamefont {M.}~\bibnamefont
  {Gell-Mann}},\ }\href {\doibase 10.1103/PhysRev.125.1067} {\bibfield
  {journal} {\bibinfo  {journal} {Phys.Rev.}\ }\textbf {\bibinfo {volume}
  {125}},\ \bibinfo {pages} {1067} (\bibinfo {year} {1962})}\BibitemShut
  {NoStop}%
\bibitem [{\citenamefont {Ne'eman}(1961)}]{Ne'eman:1961cd}%
  \BibitemOpen
  \bibfield  {author} {\bibinfo {author} {\bibfnamefont {Y.}~\bibnamefont
  {Ne'eman}},\ }\href {\doibase 10.1016/0029-5582(61)90134-1} {\bibfield
  {journal} {\bibinfo  {journal} {Nucl.Phys.}\ }\textbf {\bibinfo {volume}
  {26}},\ \bibinfo {pages} {222} (\bibinfo {year} {1961})}\BibitemShut
  {NoStop}%
\bibitem [{\citenamefont {Patrignani}\ \emph {et~al.}(2016)\citenamefont
  {Patrignani} \emph {et~al.}}]{Patrignani:2016xqp}%
  \BibitemOpen
  \bibfield  {author} {\bibinfo {author} {\bibfnamefont {C.}~\bibnamefont
  {Patrignani}} \emph {et~al.} (\bibinfo {collaboration} {Particle Data
  Group}),\ }\href {\doibase 10.1088/1674-1137/40/10/100001} {\bibfield
  {journal} {\bibinfo  {journal} {Chin. Phys.}\ }\textbf {\bibinfo {volume}
  {C40}},\ \bibinfo {pages} {100001} (\bibinfo {year} {2016})}\BibitemShut
  {NoStop}%
\bibitem [{\citenamefont {Yelton}\ \emph {et~al.}(2018)\citenamefont {Yelton}
  \emph {et~al.}}]{Yelton:2018mag}%
  \BibitemOpen
  \bibfield  {author} {\bibinfo {author} {\bibfnamefont {J.}~\bibnamefont
  {Yelton}} \emph {et~al.} (\bibinfo {collaboration} {Belle}),\ }\href@noop {}
  {\bibfield  {journal} {\bibinfo  {journal} {Submitted to: Phys. Rev. Lett.}\
  } (\bibinfo {year} {2018})},\ \Eprint {http://arxiv.org/abs/1805.09384}
  {arXiv:1805.09384 [hep-ex]} \BibitemShut {NoStop}%
\bibitem [{\citenamefont {Isgur}\ and\ \citenamefont
  {Karl}(1978)}]{Isgur:1978xj}%
  \BibitemOpen
  \bibfield  {author} {\bibinfo {author} {\bibfnamefont {N.}~\bibnamefont
  {Isgur}}\ and\ \bibinfo {author} {\bibfnamefont {G.}~\bibnamefont {Karl}},\
  }\href {\doibase 10.1103/PhysRevD.18.4187} {\bibfield  {journal} {\bibinfo
  {journal} {Phys. Rev.}\ }\textbf {\bibinfo {volume} {D18}},\ \bibinfo {pages}
  {4187} (\bibinfo {year} {1978})}\BibitemShut {NoStop}%
\bibitem [{\citenamefont {Xiao}\ and\ \citenamefont
  {Zhong}(2018)}]{Xiao:2018pwe}%
  \BibitemOpen
  \bibfield  {author} {\bibinfo {author} {\bibfnamefont {L.-Y.}\ \bibnamefont
  {Xiao}}\ and\ \bibinfo {author} {\bibfnamefont {X.-H.}\ \bibnamefont
  {Zhong}},\ }\href@noop {} {\  (\bibinfo {year} {2018})},\ \Eprint
  {http://arxiv.org/abs/1805.11285} {arXiv:1805.11285 [hep-ph]} \BibitemShut
  {NoStop}%
\bibitem [{\citenamefont {Aliev}\ \emph {et~al.}(2018)\citenamefont {Aliev},
  \citenamefont {Azizi}, \citenamefont {Sarac},\ and\ \citenamefont
  {Sundu}}]{Aliev:2018syi}%
  \BibitemOpen
  \bibfield  {author} {\bibinfo {author} {\bibfnamefont {T.~M.}\ \bibnamefont
  {Aliev}}, \bibinfo {author} {\bibfnamefont {K.}~\bibnamefont {Azizi}},
  \bibinfo {author} {\bibfnamefont {Y.}~\bibnamefont {Sarac}}, \ and\ \bibinfo
  {author} {\bibfnamefont {H.}~\bibnamefont {Sundu}},\ }\href@noop {} {\
  (\bibinfo {year} {2018})},\ \Eprint {http://arxiv.org/abs/1806.01626}
  {arXiv:1806.01626 [hep-ph]} \BibitemShut {NoStop}%
\bibitem [{\citenamefont {Polyakov}\ \emph {et~al.}(2018)\citenamefont
  {Polyakov}, \citenamefont {Son}, \citenamefont {Sun},\ and\ \citenamefont
  {Tandogan}}]{Polyakov:2018mow}%
  \BibitemOpen
  \bibfield  {author} {\bibinfo {author} {\bibfnamefont {M.~V.}\ \bibnamefont
  {Polyakov}}, \bibinfo {author} {\bibfnamefont {H.-D.}\ \bibnamefont {Son}},
  \bibinfo {author} {\bibfnamefont {B.-D.}\ \bibnamefont {Sun}}, \ and\
  \bibinfo {author} {\bibfnamefont {A.}~\bibnamefont {Tandogan}},\ }\href@noop
  {} {\  (\bibinfo {year} {2018})},\ \Eprint {http://arxiv.org/abs/1806.04427}
  {arXiv:1806.04427 [hep-ph]} \BibitemShut {NoStop}%
\bibitem [{\citenamefont {Chao}\ \emph {et~al.}(1981)\citenamefont {Chao},
  \citenamefont {Isgur},\ and\ \citenamefont {Karl}}]{Chao:1980em}%
  \BibitemOpen
  \bibfield  {author} {\bibinfo {author} {\bibfnamefont {K.-T.}\ \bibnamefont
  {Chao}}, \bibinfo {author} {\bibfnamefont {N.}~\bibnamefont {Isgur}}, \ and\
  \bibinfo {author} {\bibfnamefont {G.}~\bibnamefont {Karl}},\ }\href {\doibase
  10.1103/PhysRevD.23.155} {\bibfield  {journal} {\bibinfo  {journal} {Phys.
  Rev.}\ }\textbf {\bibinfo {volume} {D23}},\ \bibinfo {pages} {155} (\bibinfo
  {year} {1981})}\BibitemShut {NoStop}%
\bibitem [{\citenamefont {Goity}\ \emph {et~al.}(2003)\citenamefont {Goity},
  \citenamefont {Schat},\ and\ \citenamefont {Scoccola}}]{Goity:2003ab}%
  \BibitemOpen
  \bibfield  {author} {\bibinfo {author} {\bibfnamefont {J.~L.}\ \bibnamefont
  {Goity}}, \bibinfo {author} {\bibfnamefont {C.}~\bibnamefont {Schat}}, \ and\
  \bibinfo {author} {\bibfnamefont {N.~N.}\ \bibnamefont {Scoccola}},\ }\href
  {\doibase 10.1016/S0370-2693(03)00700-7} {\bibfield  {journal} {\bibinfo
  {journal} {Phys. Lett.}\ }\textbf {\bibinfo {volume} {B564}},\ \bibinfo
  {pages} {83} (\bibinfo {year} {2003})},\ \Eprint
  {http://arxiv.org/abs/hep-ph/0304167} {arXiv:hep-ph/0304167 [hep-ph]}
  \BibitemShut {NoStop}%
\bibitem [{\citenamefont {Oh}(2007)}]{Oh:2007cr}%
  \BibitemOpen
  \bibfield  {author} {\bibinfo {author} {\bibfnamefont {Y.}~\bibnamefont
  {Oh}},\ }\href {\doibase 10.1103/PhysRevD.75.074002} {\bibfield  {journal}
  {\bibinfo  {journal} {Phys. Rev.}\ }\textbf {\bibinfo {volume} {D75}},\
  \bibinfo {pages} {074002} (\bibinfo {year} {2007})},\ \Eprint
  {http://arxiv.org/abs/hep-ph/0702126} {arXiv:hep-ph/0702126 [HEP-PH]}
  \BibitemShut {NoStop}%
\bibitem [{\citenamefont {Engel}\ \emph {et~al.}(2013)\citenamefont {Engel},
  \citenamefont {Lang}, \citenamefont {Mohler},\ and\ \citenamefont
  {Schäfer}}]{Engel:2013ig}%
  \BibitemOpen
  \bibfield  {author} {\bibinfo {author} {\bibfnamefont {G.~P.}\ \bibnamefont
  {Engel}}, \bibinfo {author} {\bibfnamefont {C.~B.}\ \bibnamefont {Lang}},
  \bibinfo {author} {\bibfnamefont {D.}~\bibnamefont {Mohler}}, \ and\ \bibinfo
  {author} {\bibfnamefont {A.}~\bibnamefont {Schäfer}} (\bibinfo
  {collaboration} {BGR}),\ }\href {\doibase 10.1103/PhysRevD.87.074504}
  {\bibfield  {journal} {\bibinfo  {journal} {Phys. Rev.}\ }\textbf {\bibinfo
  {volume} {D87}},\ \bibinfo {pages} {074504} (\bibinfo {year} {2013})},\
  \Eprint {http://arxiv.org/abs/1301.4318} {arXiv:1301.4318 [hep-lat]}
  \BibitemShut {NoStop}%
\bibitem [{\citenamefont {Guo}\ \emph {et~al.}(2018)\citenamefont {Guo},
  \citenamefont {Hanhart}, \citenamefont {Meißner}, \citenamefont {Wang},
  \citenamefont {Zhao},\ and\ \citenamefont {Zou}}]{Guo:2017jvc}%
  \BibitemOpen
  \bibfield  {author} {\bibinfo {author} {\bibfnamefont {F.-K.}\ \bibnamefont
  {Guo}}, \bibinfo {author} {\bibfnamefont {C.}~\bibnamefont {Hanhart}},
  \bibinfo {author} {\bibfnamefont {U.-G.}\ \bibnamefont {Meißner}}, \bibinfo
  {author} {\bibfnamefont {Q.}~\bibnamefont {Wang}}, \bibinfo {author}
  {\bibfnamefont {Q.}~\bibnamefont {Zhao}}, \ and\ \bibinfo {author}
  {\bibfnamefont {B.-S.}\ \bibnamefont {Zou}},\ }\href {\doibase
  10.1103/RevModPhys.90.015004} {\bibfield  {journal} {\bibinfo  {journal}
  {Rev. Mod. Phys.}\ }\textbf {\bibinfo {volume} {90}},\ \bibinfo {pages}
  {015004} (\bibinfo {year} {2018})},\ \Eprint
  {http://arxiv.org/abs/1705.00141} {arXiv:1705.00141 [hep-ph]} \BibitemShut
  {NoStop}%
\bibitem [{\citenamefont {Ali}\ \emph {et~al.}(2017)\citenamefont {Ali},
  \citenamefont {Lange},\ and\ \citenamefont {Stone}}]{Ali:2017jda}%
  \BibitemOpen
  \bibfield  {author} {\bibinfo {author} {\bibfnamefont {A.}~\bibnamefont
  {Ali}}, \bibinfo {author} {\bibfnamefont {J.~S.}\ \bibnamefont {Lange}}, \
  and\ \bibinfo {author} {\bibfnamefont {S.}~\bibnamefont {Stone}},\ }\href
  {\doibase 10.1016/j.ppnp.2017.08.003} {\bibfield  {journal} {\bibinfo
  {journal} {Prog. Part. Nucl. Phys.}\ }\textbf {\bibinfo {volume} {97}},\
  \bibinfo {pages} {123} (\bibinfo {year} {2017})},\ \Eprint
  {http://arxiv.org/abs/1706.00610} {arXiv:1706.00610 [hep-ph]} \BibitemShut
  {NoStop}%
\bibitem [{\citenamefont {Olsen}\ \emph {et~al.}(2018)\citenamefont {Olsen},
  \citenamefont {Skwarnicki},\ and\ \citenamefont {Zieminska}}]{Olsen:2017bmm}%
  \BibitemOpen
  \bibfield  {author} {\bibinfo {author} {\bibfnamefont {S.~L.}\ \bibnamefont
  {Olsen}}, \bibinfo {author} {\bibfnamefont {T.}~\bibnamefont {Skwarnicki}}, \
  and\ \bibinfo {author} {\bibfnamefont {D.}~\bibnamefont {Zieminska}},\ }\href
  {\doibase 10.1103/RevModPhys.90.015003} {\bibfield  {journal} {\bibinfo
  {journal} {Rev. Mod. Phys.}\ }\textbf {\bibinfo {volume} {90}},\ \bibinfo
  {pages} {015003} (\bibinfo {year} {2018})},\ \Eprint
  {http://arxiv.org/abs/1708.04012} {arXiv:1708.04012 [hep-ph]} \BibitemShut
  {NoStop}%
\bibitem [{\citenamefont {Jido}\ \emph {et~al.}(2003)\citenamefont {Jido},
  \citenamefont {Oller}, \citenamefont {Oset}, \citenamefont {Ramos},\ and\
  \citenamefont {Meissner}}]{Jido:2003cb}%
  \BibitemOpen
  \bibfield  {author} {\bibinfo {author} {\bibfnamefont {D.}~\bibnamefont
  {Jido}}, \bibinfo {author} {\bibfnamefont {J.~A.}\ \bibnamefont {Oller}},
  \bibinfo {author} {\bibfnamefont {E.}~\bibnamefont {Oset}}, \bibinfo {author}
  {\bibfnamefont {A.}~\bibnamefont {Ramos}}, \ and\ \bibinfo {author}
  {\bibfnamefont {U.~G.}\ \bibnamefont {Meissner}},\ }\href {\doibase
  10.1016/S0375-9474(03)01598-7} {\bibfield  {journal} {\bibinfo  {journal}
  {Nucl. Phys.}\ }\textbf {\bibinfo {volume} {A725}},\ \bibinfo {pages} {181}
  (\bibinfo {year} {2003})},\ \Eprint {http://arxiv.org/abs/nucl-th/0303062}
  {arXiv:nucl-th/0303062 [nucl-th]} \BibitemShut {NoStop}%
\bibitem [{\citenamefont {Magas}\ \emph {et~al.}(2005)\citenamefont {Magas},
  \citenamefont {Oset},\ and\ \citenamefont {Ramos}}]{Magas:2005vu}%
  \BibitemOpen
  \bibfield  {author} {\bibinfo {author} {\bibfnamefont {V.~K.}\ \bibnamefont
  {Magas}}, \bibinfo {author} {\bibfnamefont {E.}~\bibnamefont {Oset}}, \ and\
  \bibinfo {author} {\bibfnamefont {A.}~\bibnamefont {Ramos}},\ }\href
  {\doibase 10.1103/PhysRevLett.95.052301} {\bibfield  {journal} {\bibinfo
  {journal} {Phys. Rev. Lett.}\ }\textbf {\bibinfo {volume} {95}},\ \bibinfo
  {pages} {052301} (\bibinfo {year} {2005})},\ \Eprint
  {http://arxiv.org/abs/hep-ph/0503043} {arXiv:hep-ph/0503043 [hep-ph]}
  \BibitemShut {NoStop}%
\bibitem [{\citenamefont {Hyodo}\ and\ \citenamefont
  {Weise}(2008)}]{Hyodo:2007jq}%
  \BibitemOpen
  \bibfield  {author} {\bibinfo {author} {\bibfnamefont {T.}~\bibnamefont
  {Hyodo}}\ and\ \bibinfo {author} {\bibfnamefont {W.}~\bibnamefont {Weise}},\
  }\href {\doibase 10.1103/PhysRevC.77.035204} {\bibfield  {journal} {\bibinfo
  {journal} {Phys. Rev.}\ }\textbf {\bibinfo {volume} {C77}},\ \bibinfo {pages}
  {035204} (\bibinfo {year} {2008})},\ \Eprint {http://arxiv.org/abs/0712.1613}
  {arXiv:0712.1613 [nucl-th]} \BibitemShut {NoStop}%
\bibitem [{\citenamefont {Guo}\ \emph {et~al.}(2006)\citenamefont {Guo},
  \citenamefont {Shen}, \citenamefont {Chiang}, \citenamefont {Ping},\ and\
  \citenamefont {Zou}}]{Guo:2006fu}%
  \BibitemOpen
  \bibfield  {author} {\bibinfo {author} {\bibfnamefont {F.-K.}\ \bibnamefont
  {Guo}}, \bibinfo {author} {\bibfnamefont {P.-N.}\ \bibnamefont {Shen}},
  \bibinfo {author} {\bibfnamefont {H.-C.}\ \bibnamefont {Chiang}}, \bibinfo
  {author} {\bibfnamefont {R.-G.}\ \bibnamefont {Ping}}, \ and\ \bibinfo
  {author} {\bibfnamefont {B.-S.}\ \bibnamefont {Zou}},\ }\href {\doibase
  10.1016/j.physletb.2006.08.064} {\bibfield  {journal} {\bibinfo  {journal}
  {Phys. Lett.}\ }\textbf {\bibinfo {volume} {B641}},\ \bibinfo {pages} {278}
  (\bibinfo {year} {2006})},\ \Eprint {http://arxiv.org/abs/hep-ph/0603072}
  {arXiv:hep-ph/0603072 [hep-ph]} \BibitemShut {NoStop}%
\bibitem [{\citenamefont {Guo}\ \emph {et~al.}(2009)\citenamefont {Guo},
  \citenamefont {Hanhart},\ and\ \citenamefont {Meissner}}]{Guo:2009ct}%
  \BibitemOpen
  \bibfield  {author} {\bibinfo {author} {\bibfnamefont {F.-K.}\ \bibnamefont
  {Guo}}, \bibinfo {author} {\bibfnamefont {C.}~\bibnamefont {Hanhart}}, \ and\
  \bibinfo {author} {\bibfnamefont {U.-G.}\ \bibnamefont {Meissner}},\ }\href
  {\doibase 10.1140/epja/i2009-10762-1} {\bibfield  {journal} {\bibinfo
  {journal} {Eur. Phys. J.}\ }\textbf {\bibinfo {volume} {A40}},\ \bibinfo
  {pages} {171} (\bibinfo {year} {2009})},\ \Eprint
  {http://arxiv.org/abs/0901.1597} {arXiv:0901.1597 [hep-ph]} \BibitemShut
  {NoStop}%
\bibitem [{\citenamefont {Sarkar}\ \emph {et~al.}(2005)\citenamefont {Sarkar},
  \citenamefont {Oset},\ and\ \citenamefont {Vicente~Vacas}}]{Sarkar:2004jh}%
  \BibitemOpen
  \bibfield  {author} {\bibinfo {author} {\bibfnamefont {S.}~\bibnamefont
  {Sarkar}}, \bibinfo {author} {\bibfnamefont {E.}~\bibnamefont {Oset}}, \ and\
  \bibinfo {author} {\bibfnamefont {M.~J.}\ \bibnamefont {Vicente~Vacas}},\
  }\href {\doibase 10.1016/j.nuclphysa.2005.01.006,
  10.1016/j.nuclphysa.2006.09.019} {\bibfield  {journal} {\bibinfo  {journal}
  {Nucl. Phys.}\ }\textbf {\bibinfo {volume} {A750}},\ \bibinfo {pages} {294}
  (\bibinfo {year} {2005})},\ \bibinfo {note} {[Erratum: Nucl.
  Phys.A780,90(2006)]},\ \Eprint {http://arxiv.org/abs/nucl-th/0407025}
  {arXiv:nucl-th/0407025 [nucl-th]} \BibitemShut {NoStop}%
\bibitem [{\citenamefont {Xu}\ \emph {et~al.}(2016)\citenamefont {Xu},
  \citenamefont {Xie}, \citenamefont {Chen},\ and\ \citenamefont
  {Jia}}]{Si-Qi:2016gmh}%
  \BibitemOpen
  \bibfield  {author} {\bibinfo {author} {\bibfnamefont {S.-Q.}\ \bibnamefont
  {Xu}}, \bibinfo {author} {\bibfnamefont {J.-J.}\ \bibnamefont {Xie}},
  \bibinfo {author} {\bibfnamefont {X.-R.}\ \bibnamefont {Chen}}, \ and\
  \bibinfo {author} {\bibfnamefont {D.-J.}\ \bibnamefont {Jia}},\ }\href
  {\doibase 10.1088/0253-6102/65/1/53} {\bibfield  {journal} {\bibinfo
  {journal} {Commun. Theor. Phys.}\ }\textbf {\bibinfo {volume} {65}},\
  \bibinfo {pages} {53} (\bibinfo {year} {2016})},\ \Eprint
  {http://arxiv.org/abs/1510.07419} {arXiv:1510.07419 [nucl-th]} \BibitemShut
  {NoStop}%
\bibitem [{\citenamefont
  {Valderrama}(2018{\natexlab{a}})}]{Valderrama:2018knt}%
  \BibitemOpen
  \bibfield  {author} {\bibinfo {author} {\bibfnamefont {M.~P.}\ \bibnamefont
  {Valderrama}},\ }\href {\doibase 10.1103/PhysRevD.98.014022} {\bibfield
  {journal} {\bibinfo  {journal} {Phys. Rev.}\ }\textbf {\bibinfo {volume}
  {D98}},\ \bibinfo {pages} {014022} (\bibinfo {year} {2018}{\natexlab{a}})},\
  \Eprint {http://arxiv.org/abs/1805.05100} {arXiv:1805.05100 [hep-ph]}
  \BibitemShut {NoStop}%
\bibitem [{\citenamefont
  {Valderrama}(2018{\natexlab{b}})}]{Valderrama:2018sap}%
  \BibitemOpen
  \bibfield  {author} {\bibinfo {author} {\bibfnamefont {M.~P.}\ \bibnamefont
  {Valderrama}},\ }\href@noop {} {\  (\bibinfo {year} {2018}{\natexlab{b}})},\
  \Eprint {http://arxiv.org/abs/1805.10584} {arXiv:1805.10584 [hep-ph]}
  \BibitemShut {NoStop}%
\bibitem [{\citenamefont {Kadyshevsky}(1968)}]{Kadyshevsky:1967rs}%
  \BibitemOpen
  \bibfield  {author} {\bibinfo {author} {\bibfnamefont {V.~G.}\ \bibnamefont
  {Kadyshevsky}},\ }\href {\doibase 10.1016/0550-3213(68)90274-5} {\bibfield
  {journal} {\bibinfo  {journal} {Nucl. Phys.}\ }\textbf {\bibinfo {volume}
  {B6}},\ \bibinfo {pages} {125} (\bibinfo {year} {1968})}\BibitemShut
  {NoStop}%
\bibitem [{\citenamefont {Guo}\ \emph {et~al.}(2014)\citenamefont {Guo},
  \citenamefont {Hidalgo-Duque}, \citenamefont {Nieves}, \citenamefont
  {Ozpineci},\ and\ \citenamefont {Valderrama}}]{Guo:2014hqa}%
  \BibitemOpen
  \bibfield  {author} {\bibinfo {author} {\bibfnamefont {F.~K.}\ \bibnamefont
  {Guo}}, \bibinfo {author} {\bibfnamefont {C.}~\bibnamefont {Hidalgo-Duque}},
  \bibinfo {author} {\bibfnamefont {J.}~\bibnamefont {Nieves}}, \bibinfo
  {author} {\bibfnamefont {A.}~\bibnamefont {Ozpineci}}, \ and\ \bibinfo
  {author} {\bibfnamefont {M.~P.}\ \bibnamefont {Valderrama}},\ }\href
  {\doibase 10.1140/epjc/s10052-014-2885-4} {\bibfield  {journal} {\bibinfo
  {journal} {Eur. Phys. J.}\ }\textbf {\bibinfo {volume} {C74}},\ \bibinfo
  {pages} {2885} (\bibinfo {year} {2014})},\ \Eprint
  {http://arxiv.org/abs/1404.1776} {arXiv:1404.1776 [hep-ph]} \BibitemShut
  {NoStop}%
\bibitem [{\citenamefont {Manohar}\ and\ \citenamefont
  {Georgi}(1984)}]{Manohar:1983md}%
  \BibitemOpen
  \bibfield  {author} {\bibinfo {author} {\bibfnamefont {A.}~\bibnamefont
  {Manohar}}\ and\ \bibinfo {author} {\bibfnamefont {H.}~\bibnamefont
  {Georgi}},\ }\href {\doibase 10.1016/0550-3213(84)90231-1} {\bibfield
  {journal} {\bibinfo  {journal} {Nucl. Phys.}\ }\textbf {\bibinfo {volume}
  {B234}},\ \bibinfo {pages} {189} (\bibinfo {year} {1984})}\BibitemShut
  {NoStop}%
\bibitem [{\citenamefont {Lu}\ \emph {et~al.}(2017)\citenamefont {Lu},
  \citenamefont {Geng},\ and\ \citenamefont {Valderrama}}]{Lu:2017dvm}%
  \BibitemOpen
  \bibfield  {author} {\bibinfo {author} {\bibfnamefont {J.-X.}\ \bibnamefont
  {Lu}}, \bibinfo {author} {\bibfnamefont {L.-S.}\ \bibnamefont {Geng}}, \ and\
  \bibinfo {author} {\bibfnamefont {M.~P.}\ \bibnamefont {Valderrama}},\
  }\href@noop {} {\  (\bibinfo {year} {2017})},\ \Eprint
  {http://arxiv.org/abs/1706.02588} {arXiv:1706.02588 [hep-ph]} \BibitemShut
  {NoStop}%
\bibitem [{\citenamefont {Haidenbauer}\ \emph {et~al.}(2017)\citenamefont
  {Haidenbauer}, \citenamefont {Petschauer}, \citenamefont {Kaiser},
  \citenamefont {Meißner},\ and\ \citenamefont {Weise}}]{Haidenbauer:2017sws}%
  \BibitemOpen
  \bibfield  {author} {\bibinfo {author} {\bibfnamefont {J.}~\bibnamefont
  {Haidenbauer}}, \bibinfo {author} {\bibfnamefont {S.}~\bibnamefont
  {Petschauer}}, \bibinfo {author} {\bibfnamefont {N.}~\bibnamefont {Kaiser}},
  \bibinfo {author} {\bibfnamefont {U.-G.}\ \bibnamefont {Meißner}}, \ and\
  \bibinfo {author} {\bibfnamefont {W.}~\bibnamefont {Weise}},\ }\href
  {\doibase 10.1140/epjc/s10052-017-5309-4} {\bibfield  {journal} {\bibinfo
  {journal} {Eur. Phys. J.}\ }\textbf {\bibinfo {volume} {C77}},\ \bibinfo
  {pages} {760} (\bibinfo {year} {2017})},\ \Eprint
  {http://arxiv.org/abs/1708.08071} {arXiv:1708.08071 [nucl-th]} \BibitemShut
  {NoStop}%
\bibitem [{\citenamefont {Albaladejo}\ \emph {et~al.}(2015)\citenamefont
  {Albaladejo}, \citenamefont {Guo}, \citenamefont {Hidalgo-Duque},
  \citenamefont {Nieves},\ and\ \citenamefont
  {Valderrama}}]{Albaladejo:2015dsa}%
  \BibitemOpen
  \bibfield  {author} {\bibinfo {author} {\bibfnamefont {M.}~\bibnamefont
  {Albaladejo}}, \bibinfo {author} {\bibfnamefont {F.~K.}\ \bibnamefont {Guo}},
  \bibinfo {author} {\bibfnamefont {C.}~\bibnamefont {Hidalgo-Duque}}, \bibinfo
  {author} {\bibfnamefont {J.}~\bibnamefont {Nieves}}, \ and\ \bibinfo {author}
  {\bibfnamefont {M.~P.}\ \bibnamefont {Valderrama}},\ }\href {\doibase
  10.1140/epjc/s10052-015-3753-6} {\bibfield  {journal} {\bibinfo  {journal}
  {Eur. Phys. J.}\ }\textbf {\bibinfo {volume} {C75}},\ \bibinfo {pages} {547}
  (\bibinfo {year} {2015})},\ \Eprint {http://arxiv.org/abs/1504.00861}
  {arXiv:1504.00861 [hep-ph]} \BibitemShut {NoStop}%
\bibitem [{\citenamefont {Pavón~Valderrama}\ and\ \citenamefont
  {Phillips}(2015)}]{Valderrama:2014vra}%
  \BibitemOpen
  \bibfield  {author} {\bibinfo {author} {\bibfnamefont {M.}~\bibnamefont
  {Pavón~Valderrama}}\ and\ \bibinfo {author} {\bibfnamefont {D.~R.}\
  \bibnamefont {Phillips}},\ }\href {\doibase 10.1103/PhysRevLett.114.082502}
  {\bibfield  {journal} {\bibinfo  {journal} {Phys. Rev. Lett.}\ }\textbf
  {\bibinfo {volume} {114}},\ \bibinfo {pages} {082502} (\bibinfo {year}
  {2015})},\ \Eprint {http://arxiv.org/abs/1407.0437} {arXiv:1407.0437
  [nucl-th]} \BibitemShut {NoStop}%
\bibitem [{\citenamefont {Valderrama}(2016)}]{Valderrama:2016koj}%
  \BibitemOpen
  \bibfield  {author} {\bibinfo {author} {\bibfnamefont {M.~P.}\ \bibnamefont
  {Valderrama}},\ }\href {\doibase 10.1142/S021830131641007X} {\bibfield
  {journal} {\bibinfo  {journal} {Int. J. Mod. Phys.}\ }\textbf {\bibinfo
  {volume} {E25}},\ \bibinfo {pages} {1641007} (\bibinfo {year} {2016})},\
  \Eprint {http://arxiv.org/abs/1604.01332} {arXiv:1604.01332 [nucl-th]}
  \BibitemShut {NoStop}%
\bibitem [{\citenamefont {Samios}\ \emph {et~al.}(1974)\citenamefont {Samios},
  \citenamefont {Goldberg},\ and\ \citenamefont {Meadows}}]{Samios:1974tw}%
  \BibitemOpen
  \bibfield  {author} {\bibinfo {author} {\bibfnamefont {N.~P.}\ \bibnamefont
  {Samios}}, \bibinfo {author} {\bibfnamefont {M.}~\bibnamefont {Goldberg}}, \
  and\ \bibinfo {author} {\bibfnamefont {B.~T.}\ \bibnamefont {Meadows}},\
  }\href {\doibase 10.1103/RevModPhys.46.49} {\bibfield  {journal} {\bibinfo
  {journal} {Rev.Mod.Phys.}\ }\textbf {\bibinfo {volume} {46}},\ \bibinfo
  {pages} {49} (\bibinfo {year} {1974})}\BibitemShut {NoStop}%
\bibitem [{\citenamefont {Lin}\ and\ \citenamefont {Zou}(2018)}]{Lin:2018nqd}%
  \BibitemOpen
  \bibfield  {author} {\bibinfo {author} {\bibfnamefont {Y.-H.}\ \bibnamefont
  {Lin}}\ and\ \bibinfo {author} {\bibfnamefont {B.-S.}\ \bibnamefont {Zou}},\
  }\href@noop {} {\  (\bibinfo {year} {2018})},\ \Eprint
  {http://arxiv.org/abs/1807.00997} {arXiv:1807.00997 [hep-ph]} \BibitemShut
  {NoStop}%
\bibitem [{\citenamefont {Ramos}\ \emph {et~al.}(2002)\citenamefont {Ramos},
  \citenamefont {Oset},\ and\ \citenamefont {Bennhold}}]{Ramos:2002xh}%
  \BibitemOpen
  \bibfield  {author} {\bibinfo {author} {\bibfnamefont {A.}~\bibnamefont
  {Ramos}}, \bibinfo {author} {\bibfnamefont {E.}~\bibnamefont {Oset}}, \ and\
  \bibinfo {author} {\bibfnamefont {C.}~\bibnamefont {Bennhold}},\ }\href
  {\doibase 10.1103/PhysRevLett.89.252001} {\bibfield  {journal} {\bibinfo
  {journal} {Phys. Rev. Lett.}\ }\textbf {\bibinfo {volume} {89}},\ \bibinfo
  {pages} {252001} (\bibinfo {year} {2002})},\ \Eprint
  {http://arxiv.org/abs/nucl-th/0204044} {arXiv:nucl-th/0204044 [nucl-th]}
  \BibitemShut {NoStop}%
\bibitem [{\citenamefont {Guzey}\ and\ \citenamefont
  {Polyakov}(2005)}]{Guzey:2005vz}%
  \BibitemOpen
  \bibfield  {author} {\bibinfo {author} {\bibfnamefont {V.}~\bibnamefont
  {Guzey}}\ and\ \bibinfo {author} {\bibfnamefont {M.~V.}\ \bibnamefont
  {Polyakov}},\ }\href@noop {} {\  (\bibinfo {year} {2005})},\ \Eprint
  {http://arxiv.org/abs/hep-ph/0512355} {arXiv:hep-ph/0512355 [hep-ph]}
  \BibitemShut {NoStop}%
\bibitem [{\citenamefont {Baru}\ \emph {et~al.}(2004)\citenamefont {Baru},
  \citenamefont {Haidenbauer}, \citenamefont {Hanhart}, \citenamefont
  {Kalashnikova},\ and\ \citenamefont {Kudryavtsev}}]{Baru:2003qq}%
  \BibitemOpen
  \bibfield  {author} {\bibinfo {author} {\bibfnamefont {V.}~\bibnamefont
  {Baru}}, \bibinfo {author} {\bibfnamefont {J.}~\bibnamefont {Haidenbauer}},
  \bibinfo {author} {\bibfnamefont {C.}~\bibnamefont {Hanhart}}, \bibinfo
  {author} {\bibfnamefont {{\relax Yu}.}~\bibnamefont {Kalashnikova}}, \ and\
  \bibinfo {author} {\bibfnamefont {A.~E.}\ \bibnamefont {Kudryavtsev}},\
  }\href {\doibase 10.1016/j.physletb.2004.01.088} {\bibfield  {journal}
  {\bibinfo  {journal} {Phys. Lett.}\ }\textbf {\bibinfo {volume} {B586}},\
  \bibinfo {pages} {53} (\bibinfo {year} {2004})},\ \Eprint
  {http://arxiv.org/abs/hep-ph/0308129} {arXiv:hep-ph/0308129 [hep-ph]}
  \BibitemShut {NoStop}%
\bibitem [{\citenamefont {Huang}\ \emph {et~al.}(2018)\citenamefont {Huang},
  \citenamefont {Liu}, \citenamefont {Lu}, \citenamefont {Xie},\ and\
  \citenamefont {Geng}}]{Huang:2018wth}%
  \BibitemOpen
  \bibfield  {author} {\bibinfo {author} {\bibfnamefont {Y.}~\bibnamefont
  {Huang}}, \bibinfo {author} {\bibfnamefont {M.-Z.}\ \bibnamefont {Liu}},
  \bibinfo {author} {\bibfnamefont {J.-X.}\ \bibnamefont {Lu}}, \bibinfo
  {author} {\bibfnamefont {J.-J.}\ \bibnamefont {Xie}}, \ and\ \bibinfo
  {author} {\bibfnamefont {L.-S.}\ \bibnamefont {Geng}},\ }\href@noop {} {\
  (\bibinfo {year} {2018})},\ \Eprint {http://arxiv.org/abs/1807.06485}
  {arXiv:1807.06485 [hep-ph]} \BibitemShut {NoStop}%
\bibitem [{\citenamefont {Pavao}\ and\ \citenamefont
  {Oset}(2018)}]{Pavao:2018xub}%
  \BibitemOpen
  \bibfield  {author} {\bibinfo {author} {\bibfnamefont {R.}~\bibnamefont
  {Pavao}}\ and\ \bibinfo {author} {\bibfnamefont {E.}~\bibnamefont {Oset}},\
  }\href@noop {} {\  (\bibinfo {year} {2018})},\ \Eprint
  {http://arxiv.org/abs/1808.01950} {arXiv:1808.01950 [hep-ph]} \BibitemShut
  {NoStop}%
\end{thebibliography}

%

\end{document}